\documentclass[twocolumn,showpacs,preprintnumbers,amsmath,amssymb,prl]{revtex4}

\usepackage{graphicx}
\usepackage{dcolumn}
\usepackage{bm}

\begin{document}


\title{
Nonequilibrium electron transport in strongly correlated molecular junctions}

\author{J. E. Han}
\affiliation{
Department of Physics, State University of New York at Buffalo, Buffalo, NY 14260, USA}

\date{\today}

\begin{abstract}

We investigate models of molecular junctions
which constitute minimal Hamiltonians to account for zero-bias-anomaly and
the satellite features of inelastic transport by
molecular phonons. Through nonlinear transport calculations
with the imaginary-time nonequilibrium formalism, 
a HOMO-LUMO model with Anderson-Holstein interaction is shown to produce
co-tunneling conductance peak in the vicinity of Kondo resonance 
which is mediated by a re-emergent
many-body resonance assisted by phonon excitations at bias equal to the
phonon frequency. Destruction of the
resonance leads to negative-differential-resistance in the sequential
tunneling regime.

\end{abstract}


\maketitle

Strong correlation in nonequilibrium electron transport 
has emerged as one of the most exciting fields of condensed
matter physics. The research in this field so far has been mostly driven by
semiconductor-fabricated quantum dots (QDs). The 
zero-bias anomaly (ZBA) phenomena have been extensively studied in the
context of Kondo phenomena~\cite{cronenwett,grobis}. In recent years,
similar ZBA phenomenon in molecular junctions~\cite{yu,yu2,osorio,scott} 
has generated tremendous excitement
for possible different mechanisms for strongly correlated transport.

Currently, the research on molecular junctions in strongly correlated
regime is, both experimentally and theoretically, at an early stage and
little is known for the underlying transport mechanisms. One of the most
outstanding transport phenomena in molecular devices is the co-existence
of the ZBA and the inelastic conductance peaks, presumably due to
molecular phonons~\cite{yu,yu2,osorio}. Theoretically, the
strong correlation in molecular systems poses a great challenge since
strong Coulomb and electron-phonon (el-ph) interactions make
perturbative approaches unreliable. Only recently, strong correlation
physics in the Anderson-Holstein model has been understood for
equilibrium systems\cite{c60,hewson,cornaglia}. Most works on
nonequilibrium transport in molecular systems have been
perturbative and often excluded Coulomb
interaction~\cite{mitra,flensberg}.  Although nonperturbative
nonequilibrium theories have seen important
breakthroughs~\cite{hershfield,imaginary,anders,rosch,mehta} 
in the past few years, the methods
have not been adequate to tackle complex models such as molecular
junctions. 

The main goal of this work is to identify minimal Anderson-Holstein
models which can describe the Kondo anomaly and the inelastic features at
finite source-drain bias, and reproduce some of experimental
findings~\cite{yu,yu2,scott,osorio}. We apply the recently developed
imaginary-time theory~\cite{imaginary} and numerically solve the
Anderson-Holstein models via quantum Monte Carlo (QMC)
method~\cite{hirsch}.  Due to the diverse molecular systems, it is very
important at this stage to have guiding principles to categorize
molecular models for different transport phenomena. The main system of
focus here are molecular quantum dots which exhibit the ZBA accompanied
by conductance oscillations at bias near the ZBA energy scale, which have
been often attributed to the molecular vibrations. This problem, from
the strong-correlation point of view, is quite puzzling since near the
Kondo anomaly the charge fluctuations are strongly suppressed and phonons
which interact with electric charge fluctuations are effectively
decoupled~\cite{c60,sangiovanni}. In single-orbital Anderson-Holstein
models, it has been shown that phonon spectral features in strong
Coulomb limit are weak.

\begin{figure}[bt]
\rotatebox{0}{\resizebox{2.5in}{!}{\includegraphics{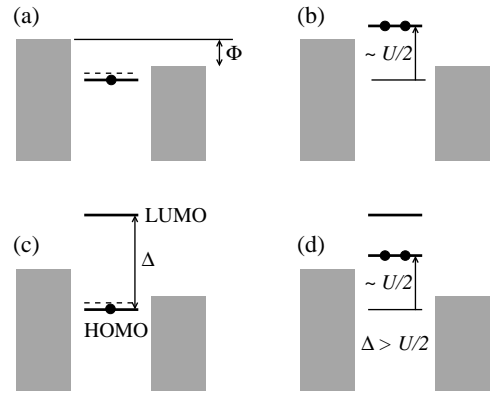}}}
\caption{
Schematic energy diagrams for isolated molecular configurations with
respect to source/drain reservoirs under bias $\Phi$.
(a) Single-orbital model with one electron occupying the level at energy
$\epsilon_d$. Phonon excitation level is marked by a dashed line. (b)
With an extra electron,
the energy level is pushed up $\epsilon_d+U\sim U/2$
by Coulomb repulsion. (c) HOMO-LUMO model
with the level spacing $\Delta$. (d) Charge-excited state. With
$\Delta\gg U/2$, the LUMO level is nearly empty.
}
\label{fig1}\end{figure}

To resolve the issue, we consider two scenarios. First, we note
that, at finite bias, the strong correlation effects become weaker, as
shown in the disappearance of Kondo peak~\cite{cronenwett,grobis}. Then
incoherent charge fluctuations induced by nonequilibrium may enhance the
effective el-ph interaction.  Within this scenario, we study the
single-orbital (SO) Anderson-Holstein model. [See FIG.~\ref{fig1}(a-b)]
Second, we study a two-orbital model with
highest-occupied-molecular-orbital (HOMO) and
lowest-unoccupied-molecular-orbital (LUMO). Despite being more
realistic, HOMO/LUMO (HL) models have not been extensively studied
due to their complexity.  Multiple orbitals allow the electron
density distortion to couple to molecular distortions, \textit{i.e.}
molecular Jahn-Teller (JT) modes, without invoking on-site charge
fluctuations. It has been shown that in strong correlation
limit~\cite{c60}, the JT coupling becomes very effective. [See
FIG.~\ref{fig1}(c-d)]

For both models, the electron source/drain reservoirs are modeled by the
Hamiltonian $H_c=\sum_{k\alpha\sigma}\epsilon_{\alpha
k\sigma}c^\dagger_{\alpha k\sigma}c_{\alpha k\sigma}$ in terms of the
electron creation (annihilation) operator $c^\dagger_{\alpha
k\sigma}(c_{\alpha k\sigma})$ with $k$ the continuum index, $\sigma$
spin index, and the reservoir index $\alpha=\pm 1$ for source ($L$) and drain ($R$),
respectively. The charge part of the QD Hamiltonian for the
single-orbital model is
\begin{equation}
H_{ch,SO}=\epsilon_d\sum_\sigma
n_{d\sigma}+\frac{U}{2}(n_{d\uparrow}+n_{d\downarrow}-1)^2,
\end{equation}
with the number operator $n_{d\sigma}=d^\dagger_\sigma d_\sigma$ of the QD
orbital $d^\dagger_\sigma$, the level
energy $\epsilon_d$ and the Coulomb parameter $U$.
The Holstein phonon and the el-ph coupling for the SO model can be written as
\begin{equation}
H_{ph,SO}=\omega_{ph}a^\dagger a
+g_{ep}(a^\dagger+a)(n_{d\uparrow}+n_{d\downarrow}-1),
\end{equation}
with $a^\dagger$ for creation of phonon, $\omega_{ph}$ the phonon frequency,
$g_{ep}$ the el-ph coupling constant.
The tunneling part is 
$H_{t,SO}=-t\sum_{\alpha k\sigma}(d^\dagger_\sigma c_{\alpha
k\sigma}+h.c.)$ with the hopping parameter $t$. 
Here the tunneling rate is parametrized
by the hybridization function $\Gamma_{L,R}=\pi t^2 N(0)$ with $N(0)$ the
density of state of the reservoirs. Throughout this work, we assume
$\Gamma_L=\Gamma_R$ and use $\Gamma=\Gamma_L+\Gamma_R=1$ as the unit of
energy. The total Hamiltonian is $H=H_c+H_t+H_{ch}+H_{ph}$. 
Here, we study the regime where the phonon frequency is 
comparable to the Kondo temperature and we set 
$\omega_{ph}\sim\Gamma$, in contrast to
semiconductor QD models where the phonon-excited QD-levels are
discrete and well-defined ($\omega_{ph}>\Gamma$)~\cite{lake}.

We solve steady-state nonequilibrium using the imaginary-time
formalism~\cite{imaginary}. This method combines the nonequilibrium
quantum statistics and quantum dynamics within the equilibrium theory
via an imaginary-time Hamiltonian with complex chemical
potentials parametrized by the
\textit{Matsubara voltage} $\varphi_m=4\pi mT$ as
\begin{equation}
\hat{K}(i\varphi_m)=
\hat{K}_0(i\varphi_m)+\hat{V}
=\hat{H}_0+\frac12(i\varphi_m-\Phi)\hat{Y}_0+\hat{V},
\end{equation}
where the many-body interaction is given by $\hat{V}$ and the
non-interacting part by $\hat{H}_0=H_c+H_t+\epsilon_d\sum_\sigma
n_{d\sigma}$. Population of the
scattering states for source and drain in the non-interacting limit is
imposed by the operator~\cite{hershfield,han}
$
\hat{Y}_0=\sum_{k\sigma}(
\psi^\dagger_{L k \sigma}\psi_{L k \sigma}
-\psi^\dagger_{R k \sigma}\psi_{R k \sigma}),
$
with the scattering state operator $\psi^\dagger_{\alpha k \sigma}$ from
the $\alpha$-reservoir~\cite{han}.
$\hat{Y}_0$ can be exactly solved in the
non-interacting limit. In a perturbation expansion with $\hat{V}$,
the quantum statistics is unaffected by $i\varphi_m$ due to
$
e^{-\beta K_0}=e^{-\beta (H_0-\frac{\Phi}{2}Y_0)},
$
since $\frac12 \beta\varphi_m \hat{Y}_0$ represents
$2\pi\times(\mbox{integer})$ with respect to
the unperturbed scattering-state basis. The quantum dynamics,
represented by an energy denominator in Green functions, is
recovered by the analytic continuation $i\varphi_m\to\Phi$. This formalism
can be shown to be equivalent to the retarded Green function in the
Keldysh formalism.

With this formalism, the equilibrium auxiliary-field QMC
method~\cite{hirsch} can be immediately applied with $K(i\varphi_m)$ as
the Hamiltonian.
The resulting
self-energy $\Sigma(i\omega_n,i\varphi_m)$ at the fermion Matsubara
frequency $\omega_n=(2n+1)\pi T$ should be analytically continued
numerically. This is achieved by making an ansatz on the spectral
representation~\cite{imaginary},
\begin{equation}
\Sigma(i\omega_n,i\varphi_m)=a(i\varphi_m)+
\sum_\gamma\int
\frac{\sigma_\gamma(\epsilon)d\epsilon}{i\omega_n+
\frac{\gamma}{2}(i\varphi_m-\Phi)-\epsilon},
\label{fit}
\end{equation}
with the spectral function $\sigma_\gamma(\epsilon)$.
The index $\gamma$ of odd integer is a combination of reservoir
indices $\alpha$ in particle-hole lines in a self-energy
diagram. This representation
is exact in the equilibrium limit
or in the second order perturbation in nonequilibrium. We use
$\sigma_\gamma(\epsilon)$ as fitting parameters to the numerical
self-energy. In the
particle-hole symmetric case, the $\omega_n$-independent term $a(i\varphi_m)$
is zero. With particle-hole asymmetry, we use a simple-pole
approximation
$
a(i\varphi_m)=a_0+a_1[(i\varphi_m-z)^{-1}
+(-i\varphi_m-z^*)^{-1}],
$
with fitting parameter $a_0,a_1,z$ with ${\rm Im}\,z<0$ for $\varphi_m>0$.
$a(i\varphi_m)$ has a weak dependence on $\varphi_m$ and the analytic
continuation has been insensitive to the choice of a fitting form.
Once all the fitting parameters are found, we set $i\varphi_m\to \Phi$
and $i\omega_n\to\omega+i\eta$, and
obtain the retarded self-energy and QD Green function $G^{ret}(\omega)$. 
The current is calculated from
\begin{equation}
I=\frac{2e}{h}\int d\omega \pi\Gamma A(\omega)
\left[f\left(\omega-\frac{\Phi}{2}\right)
-f\left(\omega+\frac{\Phi}{2}\right)\right],
\label{meir}
\end{equation}
with $A(\omega)=-\pi^{-1}G^{ret}(\omega)$~\cite{wingreen}.

\begin{figure}[bt]
\rotatebox{0}{\resizebox{3.2in}{!}{\includegraphics{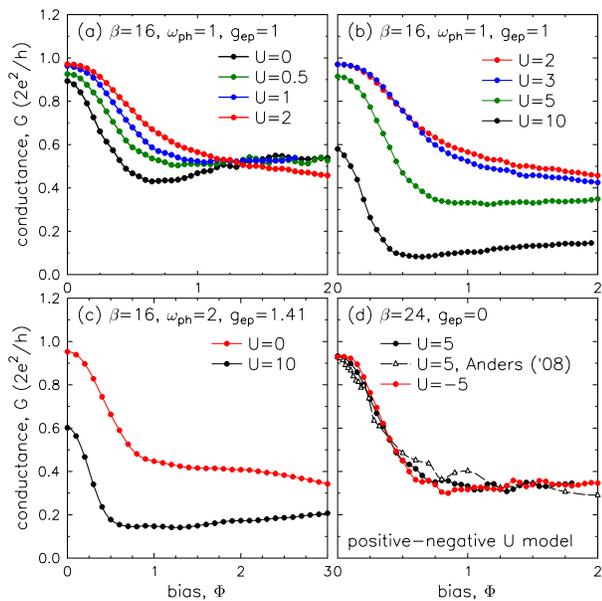}}}
\caption{(Color online)
Differential conductance in single-orbital Anderson-Holstein model.
(a) Conductance with large el-ph coupling
$(2g_{ep}^2/\omega_{ph}>U)$ in the charge-Kondo regime. As $U$
increases from zero, correlation effects become weaker.
(b) Spin-Kondo regime recovered with large $U$. 
(c) Larger phonon frequency at fixed $g_{ep}^2/\omega_{ph}=1$.
(d) Comparison of the
$g_{ep}=0$ limit to Ref~\cite{anders} and the negative $U$ model.
The unit of energy is the non-interacting line-broadening $\Gamma$.}
\label{fig2}\end{figure}

The differential conductance $G=e(dI/d\Phi)$ for the SO model is shown
in FIG.~\ref{fig2} in the particle-hole symmetric limit
$\epsilon_d=-U/2$. At $U=0$, the system has a ZBA peak with the HWHM
($\Phi_{HWHM}\sim 0.3$), much reduced from the non-interacting value
$\Phi^0_{HWHM}=2$, demonstrating the charge-Kondo
effect~\cite{cornaglia,coleman}. As $U$ approaches the el-ph binding
energy $2g_{ep}^2/\omega_{ph}$, the correlation effect becomes weaker
indicating the competition of the attractive el-ph and repulsive Coulomb
interactions. As $U$ grows further, the system approaches the usual
spin-Kondo regime.  The fine structure in the conductance shows faint
oscillations, reminiscent of some of the experiments~\cite{osorio}. However,
given the numerical uncertainties, we cannot conclude that the features
are a direct manifestation of inelastic excitation of phonon quanta. In
an extensive set of calculations we found no well-defined phonon
satellites near the ZBA energy scale. To further support the idea, we
doubled the phonon frequency ($\omega_{ph}=2$) at fixed
$g_{ep}^2/\omega_{ph}$. FIG.~\ref{fig2}(c) shows similar results as
(a-b) with weaker fine structures, possibly from more efficient QMC
sampling at high $\omega_{ph}$. We conclude that the main effect of
el-ph interaction in the SO model is the reduction of the Coulomb
interaction and that the conductance at high bias is more related to the
el-ph scattering than to the energy exchange with phonon.

FIG.~\ref{fig2}(d) shows results from pure Anderson models.  In the
previous work of Han and Heary~\cite{imaginary}, the particle-hole
symmetry condition on the spectral function in Eq.~(\ref{fit}),
$\sigma_\gamma(\epsilon)=\sigma_{-\gamma}(-\epsilon)$, was not properly
imposed and they obtained underestimated ZBA peak-widths. A modified fit
gives an improved agreement with Ref.~\cite{anders}. We also test the
idea whether the nonequilibrium imposed on the charge variable as
opposed to the spin variable has any significant effects on the
nonlinear transport. The positive-negative $U$ models are
interchangeable in equilibrium~\cite{coleman} by switching the charge
and spin variables.  The calculation shows that the difference between
the two models even at high bias is minimal.

We now turn to the HOMO/LUMO model. We denote the QD levels by $d^\dagger_{i\sigma}$
with $i=1,2$ for HOMO and LUMO, respectively. The charge part of the
Hamiltonian is 
\begin{equation}
H_{ch,HL}=\sum_{\sigma,i=1,2}\epsilon_{i}
n_{i\sigma}+\frac{U}{2}(n_{1\uparrow}+n_{1\downarrow}-1)^2,
\end{equation}
with the HOMO level at $\epsilon_1=\epsilon_d$, the LUMO level at
$\epsilon_2=\epsilon_d+\Delta$. The Coulomb interaction is set to act only on the
HOMO level, since the inclusion of the LUMO led to severe sign-problems
in QMC calculations. However, in the following calculations, the
negligence of Coulomb interaction on the LUMO becomes a
reasonable approximation since we choose the level
spacing $\Delta$ much larger than the charging energy 
($\Delta\gg\frac{U}{2}$) such that the LUMO level is mostly empty. (See
FIG.~\ref{fig1})

We model the el-ph coupling via the Jahn-Teller phonons
of $a^\dagger_m$ $(m=1,2)$ as
\begin{equation}
H_{ph,HL}=\omega_{ph}\sum_m a^\dagger_m a_m
+\sum_{ijm\sigma}(a^\dagger_m+a_m)d^\dagger_{i\sigma}V^{(m)}_{ij}d_{j\sigma},
\end{equation}
with the $2\times 2$ JT coupling matrix~\cite{manini} given as
$
V^{(1)}=g_{ep}\hat{\sigma}_z,
V^{(2)}=g_{ep}\hat{\sigma}_x,
$
with the Pauli matrices $\hat{\sigma}_z$ and $\hat{\sigma}_x$.
The second phonon ($m=2$) makes direct electronic transitions between
the LUMO and HOMO without tunneling through reservoirs. The
tunneling part is given as $H_{t,HL}=-t\sum_{\alpha k\sigma}\sum_i
(d^\dagger_{i\sigma}c_{\alpha k\sigma}+h.c)$. The average sign in the
QMC calculations has been moderate at $Sign=0.5-1.0$.

\begin{figure}[bt]
\rotatebox{0}{\resizebox{3.2in}{!}{\includegraphics{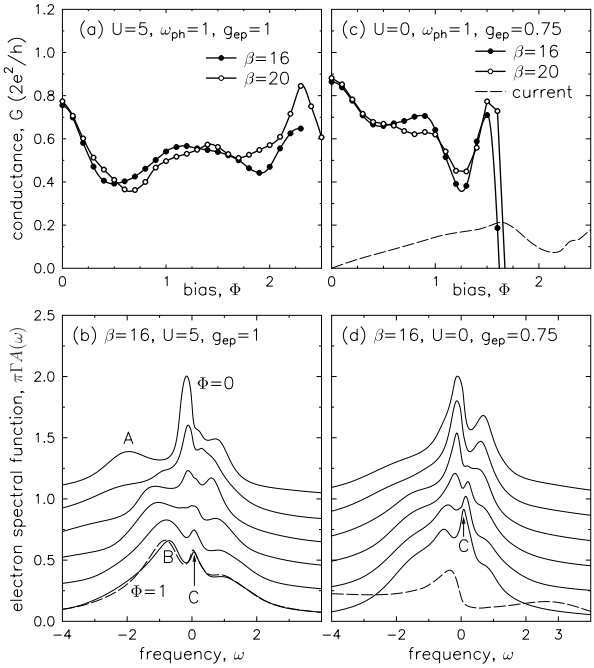}}}
\caption{
Differential conductance in the HOMO-LUMO model with
Anderson-(Jahn-Teller) Holstein model.
(a) Conductance peaks for the Kondo anomaly and inelastic
co-tunneling at $\Phi=\omega_{ph}$.
(b) Spectral functions for bias $\Phi=0,\cdots,1$ with
the interval $0.2$. Curves are shifted for clarity. With increasing
$\Phi$, the Kondo peak disappears and the Coulomb peak (peak $A$) shifts to phonon
excitation energy (peak $B$). As $\Phi\to\omega_{ph}$, new resonance
(peak $C$) emerges inside
the voltage window and contributes to the co-tunneling conductance peak at
$\Phi\approx\omega_{ph}$. Dashed line is for $\beta=20$. (c) Conductance in pure
el-ph limit. A strong
negative-differential-resistance (NDR) effect appears at $\Phi/2=\omega_{ph}$
in the sequential tunneling regime.
(d) Disappearance of the resonance (dashed line at
$\Phi/2=\omega_{ph}$) leads to the NDR.
}
\label{fig3}\end{figure}

The conductance in FIG.~\ref{fig3} shows clear phonon excitation
peaks at $\Phi=\omega_{ph}$. The HOMO-LUMO level spacing is set at
$\Delta=15$, much greater than any other energy scales. With the HOMO level
at $\epsilon_d=-1.1$, the HOMO occupation number $n_{HOMO}\sim 0.57$
at $\Phi=0$ for (a). For (c), $\epsilon_d=-1.5$ and $n_{HOMO}\sim 0.48$. The main
features of the conductance are the ZBA and the peak at
$\Phi\approx\omega_{ph}$.

To understand how the co-tunneling via phonon excitation arises,
we study the QD spectral function as
$\Phi\to\omega_{ph}$.  In FIG.~\ref{fig2}(b), spectral functions are
plotted for $\Phi=0,\cdots,1$
with the interval of $0.2$. Curves are off-set for clarity.
Destruction of the ZBA resonance is similar to the pure Anderson
models~\cite{imaginary,anders}. At $\Phi=0$, the other dominant peak is
the charge excitation peak, marked by $A$ in (b). As $\Phi$ grows, the
peak $A$ quickly migrates to the peak $B$ at $\omega=\omega_{ph}$ for
phonon excitation. The peak $B$ 
is not directly responsible for the conductance
peak at $\Phi=\omega_{ph}$ since the peak $B$ is outside the transport energy window
$[-\Phi/2,\Phi/2]$ in Eq.~(\ref{meir}).

The co-tunneling transport is carried by a new emerging resonance as indicated by
peak $C$ in FIG.~\ref{fig3}(b). As the ZBA peak disappears, another resonance
peak inside the transport energy window becomes stronger.
At $\Phi=\omega_{ph}$, the mismatch of electron Fermi
energies from the source and drain is compensated by an emission of a
phonon quantum. Then effectively the same electronic chemical
potentials on both reservoirs seem to result in a Kondo-like phonon-assisted 
many-body resonance.

\begin{figure}[bt]
\rotatebox{0}{\resizebox{3.2in}{!}{\includegraphics{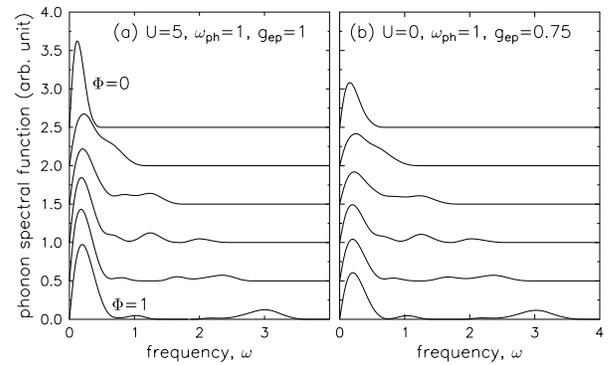}}}
\caption{
(a) Phonon spectral function of HOMO/LUMO model for parameters of FIG.~\ref{fig3}(a).
At $\Phi=0$, phonon is strongly renormalized. With increasing $\Phi$,
higher phonon quanta are created by absorbing the chemical potential
difference in the electron reservoirs.
(b) Similar results for FIG.~\ref{fig3}(b).
}
\label{fig4}\end{figure}

In a pure phonon model in FIG.~\ref{fig3}(c-d), the conductance behavior remains
qualitatively the same, but it showed a strong
negative-differential-resistance (NDR) behavior near
$\Phi/2\approx\omega_{ph}$ in the inelastic sequential tunneling regime.
As shown in a dashed line at $\Phi=2\omega_{ph}$, the spectral weight
shifts to high frequency and the resonance peak $C$ is destroyed, which
leads to the NDR.  Similar polaronic effects to NDR in molecular systems
have been reported previously~\cite{galperin}. Although this behavior is
robustly reproduced at different parameters, it should be
mentioned that this is the regime where the ansatz, Eq.~(\ref{fit}), starts
to deviate from the QMC data significantly and
vertex corrections may be necessary.

Finally, nonequilibrium-induced multi-phonon modes are shown
in FIG.~\ref{fig4}. The phonon spectral functions are calculated by the
same ansatz, Eq.~(\ref{fit}), but with even integers $\gamma$.
At $\Phi=0$, the phonon frequency is highly
renormalized from $\omega_{ph}$, but phonons are mostly in the
lowest quantum state. As $\Phi$ increases, the spectral weight transfers to
multiple phonon modes as previously predicted~\cite{mitra,flensberg}.

I thank useful discussions with F. Anders, H. van der Zant, R. Heary.
I acknowledge support from the National Science Foundation
DMR-0426826 and computing resources at CCR of SUNY
Buffalo.

\end{document}